\documentclass[11pt]{emulateapj}\usepackage{times}
\usepackage{amssymb}

\newcommand{\TO}{\Delta t_{\rm o}}  
\newcommand{\TS}{\Delta t_{\rm s}}
\newcommand{\TB}{\Delta t_{\rm b}} 
\newcommand{\SP}{\{\TO,\TS,\TB\}}
\newcommand{\PO}{P_{\rm o}(\nu)}  
\newcommand{\PM}{P_{\rm}(\nu)}
\newcommand{\sinc}{\mathrm{sinc}}
\newcommand{\msun}{\hbox{$\rm\thinspace M_{\odot}$}}
\newcommand{\mbh}{{\thinspace M_{\rm BH}}}
\newcommand{\rxte}    {\emph{RXTE}} 
 
\newcommand{\xmm}     {\emph{XMM-Newton}}
\newcommand{\chandra} {\emph{Chandra}} 
\newcommand{\asca}    {\emph{ASCA}}
\newcommand{\rms}{\sigma_{\rm rms}}
\newcommand{\rmssq}{\sigma^2_{\rm rms}}

\begin{document}

\shorttitle{\textsc{Sampling Effects in the $\rmssq$-$\mbh$
    Correlation}} 
\shortauthors{\textsc{PESSAH}}

\title{\textsc{MASS MEASUREMENTS OF AGN FROM MULTI-LORENTZIAN MODELS
    OF X-RAY VARIABILITY. I. \\ SAMPLING EFFECTS IN THEORETICAL MODELS
    OF THE $\rmssq$-$\mbh$ CORRELATION}}

\author{Martin E. Pessah} \affil{Astronomy and Physics Departments,
University of Arizona, 933 N. Cherry Ave. Tucson, AZ, 85721;
mpessah@as.arizona.edu}

\begin{abstract}
  Recent X-ray variability studies on a large sample of Active
  Galactic Nuclei (AGN) suggest that the $log$ of the square of the
  fractional rms variability amplitude, $\rmssq$, seems to correlate
  with the $log$ of the black-hole mass, $\mbh$, with larger black
  holes being less variable for a fixed time interval.  The fact that
  the rms amplitude can be easily obtained from the observed X-ray
  light curves has motivated the theoretical modeling of the
  $\rmssq$-$\mbh$ correlation with the aim of constraining AGN masses
  based on X-ray variability.  A viable approach to addressing this
  problem is to assume an underlying power spectral density with a
  suitable mass dependence, derive the functional form of the
  $\rmssq$-$\mbh$ correlation for a given sampling pattern, and
  investigate whether the result is consistent with the observations.
  Inspired by the similarities shared by the timing properties of AGN
  and X-ray binaries, previous studies have explored model power
  spectral densities characterized by broken power laws.  For
  simplicity these studies have in general ignored the distorting
  effects that the particular sampling pattern used to obtain the
  observations imprints in the observed power spectral density.  With
  the advent of $\rxte$, however, it has been shown that the
  band-limited noise spectra of X-ray binaries are not broken power
  laws but can often be described in terms of a small set of broad
  Lorentzians, with different amplitudes and centroid frequencies.
  Motivated by the latest timing results from X-ray binaries, we
  propose that AGN broad-band noise spectra consist of a small number
  of Lorentzian components.  This assumption allows, for the first
  time, to fully account for sampling effects in theoretical models of
  X-ray variability in an analytic manner.  We show that, neglecting
  sampling effects when deriving the fractional rms from the model
  power spectral density can lead to underestimating it by a factor of
  up to $80\%$ with respect to its true value for the typical sampling
  patterns used to monitor AGN.  We discuss the implications of our
  results for the derivation of AGN masses using theoretical models of
  the $\rmssq$-$\mbh$ correlation.
\end{abstract}

\keywords{accretion disks --- black hole physics ---
galaxies:active, nuclei --- X-rays:binaries, galaxies}

\section{\textsc{Introduction}}  
\label{sec:introduction}

Most of the known accreting compact objects emitting in the X-ray
band, both in our and other galaxies, show large variations in their
fluxes (up to a factor of a few or even more in some cases) over
several decades in frequency.  X-ray binaries, hosting neutron stars
and black holes with masses between $1 \msun$ and $10 \msun$, are
known to be variable on time scales ranging from milliseconds up to
several days \citep[see][for a recent review]{VDK05}. The X-ray
emission from Active Galactic Nuclei (AGN), on the other end of the
mass spectrum, with masses in the range $10^6 \msun$ to $10^9 \msun$,
fluctuates on time scales ranging from a few minutes up to a few years
\citep[e.g.,][]{MWP81, Mush93}.  The origin of this strong variability
is currently not understood.

From the theoretical point of view, the main goal of variability
studies is to understand the physical processes that modulate the
emission of high-energy radiation and, in particular, how the
fundamental properties of the central compact object (mass, spin,
surface or lack thereof) determine the spectrum of observed
frequencies. From the phenomenological point of view, observations
provide a powerful tool for constraining mathematical models which, in
turn, provide a guide in the investigation of physical models that can
account for the observed variability.  The studies carried out over
the last decade with the \emph{Advanced Satellite for Cosmology and
Astrophysics} (\asca) and the \emph{Rossi X-ray Timing Explorer}
(\rxte) and, more recently, with the \emph{Chandra X-ray Observatory}
and \xmm\ have revolutionized the timing phenomenology of both X-ray
binaries and AGN.

In the case of galactic sources, because of the typical timescales
involved and thanks to the unprecedented timing capabilities of \rxte,
it has become possible to derive the power spectral density (hereafter
the power spectrum) of a large number of accreting binaries with
exquisite detail \citep[see e.g.,][]{VDK05}.  This made possible an
important step forward toward understanding the similarities and
differences exhibited in the power spectra of neutron stars and black
holes \citep{Miyamotoetal94,VDK94,SR00,BPK02}

Recently, it has become evident that the most salient timing features
present in many X-ray binaries can be well described by a small number
of Lorentzians with different centroid frequencies, widths, and
amplitudes \citep{Now00, BPK02}.  This phenomenological framework has
provided a solid ground for studying a number of tight correlations
exhibited by the parameters characterizing the different Lorentzian
components.  Surprisingly, these correlations are not only present in
any given source but are also maintained across sources (both neutron
stars and black holes) over several decades in frequency space
\citep{WK99, PBK99, BPK02}. Some of these correlations might even be
present in white-dwarf systems, extending the range of their validity
to lower frequencies by two orders of magnitude \citep{WW02, M02}.
The study, and eventual understanding, of these correlations offers
one of the most promising avenues for constraining theoretical models
of X-ray variability in accreting binaries.

The lower observed fluxes and longer timescales associated with AGN
have not made possible a comparable progress in our understanding of
their timing properties. Both of these aspects affect the
characterization of the power spectrum from the data at high and low
frequencies, but also, through sampling effects, within the accessible
frequency range.  If there is significant variability below or above
the lowest and/or highest frequencies probed by the observations, then
the estimation of the power spectrum from the data suffers from the
effects of red noise leakage and aliasing \citep[see,
e.g.,][]{Deeming75, Priestley89, VDK89, Press92}.

In the last few years, however, it has become clear that AGN power
spectra do present characteristic frequencies (``breaks'') on the
scale of hours to months \citep{EN99, UMP02, MEV03}. These frequencies
have been linked to similar characteristic frequencies detected in
early observations of a number of galactic black-hole candidates
\citep[see, e.g.,][]{BH90}.  Moreover, there is a growing body of
evidence suggesting that the similarities shared by the X-ray
variability of AGN and X-ray binaries extends beyond the overall shape
of their power spectra. These similarities encompass a strong linear
correlation between the rms variability amplitude and the X-ray flux,
time scale-dependent lags between the hard and soft X-ray bands, and a
similar energy-dependence of the power spectrum shape with higher
energies showing flatter slopes above the frequency break \citep[][and
references therein]{UM04}.  The remarkable similarities shared by the
global timing properties of objects with masses that differ so widely
and the fact that the characteristic dynamical frequencies observed in
AGN seem to correlate with mass, has encouraged the use of X-ray
variability as a mean for estimating supermassive black-holes masses.
Two different approaches are in use.

For the handful of AGN for which there are good available data
covering a long period of time, it is possible to identify one (and in
a few cases two) characteristic frequency(ies) in their power
spectrum.  Assuming that the same characteristic frequency can be
identified in a galactic black hole of known mass (usually Cyg X-1)
and that the characteristic break frequency scales inversely with
mass, the mass of several AGN can be inferred \citep{UMP02, MEV03,
MPU04, MGU05}.

As it is most often the case, however, the sparsity of the available
data does not permit a reliable estimation of the power spectrum.  In
these cases, it is still possible to quantify the variability directly
from the observed light curves by calculating the fractional rms
variability amplitude (or simply fractional rms), $\rms$, i.e., the
square root of the variance of the light curve normalized by the mean
flux over the period of observation after subtracting the experimental
noise \citep[see, e.g.,][]{NGM97}.  Obtaining variances from X-ray
light curves is a much less intensive observational task than
obtaining reliable power spectra. As a result, the fractional rms
variability has been measured for a relatively large sample of AGN for
a variety of sampling patterns with \asca, \rxte, and, more recently,
with \chandra\ and \xmm.

Possible correlations between the square of the fractional rms and the
X-ray luminosity have been studied in the past \citep{NGM97, TGN99,
ME01}.  More recently, the availability of mass estimates for a large
number of AGN \citep[see][for a recent compilation of $\simeq 300$
masses]{WU02} has allowed the study of correlations between $\rmssq$
and black-hole mass, $\mbh$ \citep{LY01, BZ03, ME04, ONP05}. The fact
that larger black holes are less variable for a given observation
period suggests that the observed (anti)correlation is partially due
to a corresponding difference in the size of the X-ray emitting
region.

The fact that the fractional rms can be easily obtained from the
observed X-ray light curves, and that the number of AGN with measured
variances is rapidly increasing, has motivated the theoretical
modeling of the $\rmssq$-$\mbh$ correlation.  A fruitful approach to
addressing this problem is to assume a functional form for the
underlying power spectrum that depends on mass, $P(\nu;\mbh)$, and
calculate the square of the fractional rms that one would obtain when
observing according to some sampling pattern, i.e., $\rmssq(\TO, \TS,
\TB, M_{\rm BH})$. Here, the set of values $\SP$ stands for the
duration of the observation, the sampling interval, and the binning
interval respectively.

The connection between the rms variability and the underlying power
spectrum is crucial in theoretical models of the $\rmssq$-$\mbh$
correlation.  As a first approximation, previous works \citep[][but
see also O'Neill et al. 2005]{P04, NPC04} have assumed that the
relationship between the observed fractional rms and the underlying
(model) power spectrum is given by
\begin{equation}
\label{eq:rms_nosampling}
\rmssq(\TO, \TS, \TB, M_{\rm BH}) = 2 \int_{1/\TO}^{1/2\TS} \!\! P(\nu;
M_{\rm BH}) ~d\nu ~,
\end{equation}
where sampling effects are considered on the right hand side only
through the finite limits of integration. To our knowledge, however, a
systematic study addressing how good this approximation is for the
typical sampling patterns employed in AGN observations has not been
carried out yet.

In this paper, motivated by current detailed timing studies of X-ray
binaries, we consider model AGN band-limited noise spectra that can be
described as a sum of Lorentzian components.  This framework exposes
(and allows to explore further) the underlying assumption often made
when measuring AGN masses by comparing their power spectra with those
of X-ray binaries, i.e., that the global timing properties of both
systems simply ``scale'' with mass.  An appealing feature of this
framework is that it allows an analytical approach to account for the
effects that the sampling pattern imprints on the observed power
spectrum, facilitating the inclusion of sampling effects in
theoretical models of the $\rmssq$-$\mbh$ correlation.

The paper is organized as follows.  In \S \ref{sec:sampling} we
discuss, from first principles, how the underlying power spectrum,
whatever its functional form may be, differs from the observed power
spectrum due to the finite and discrete nature of the observations.
In \S \ref{sec:lorentzian models for AGN} we motivate the use of
Lorentzians to model the broad-band noise spectra of AGN and find an
analytical expression to account for sampling effects in this type of
models.  In \S \ref{sec:contrast} we demonstrate the excellent
agreement between our analytical results and Monte Carlo simulations
designed to account for the distortions suffered by the underlying
power spectral density due to sampling effects.  Finally, in \S
\ref{sec:discussion} we discuss the potential implications of
neglecting sampling effects in theoretical models when deriving masses
from the $\rmssq$-$\mbh$ correlation.  For convenience, some of the
mathematical details are presented in the appendices.

While writing this paper, we became aware of a calculation similar to
the one that we present in \S \ref{sec:sampling} carried out by
\citet{ONP05}.  Their derivation accounts for sampling effects
directly in the power spectrum as they would be present due to
discrete sampling in a ``continuous'' monitoring campaign (i.e.,
$\TS=\TB$).  The derivation that we present here is more general and
shows explicitly how the different processes which one subjects the
light curve to (i.e., binning, sampling, and segmentation), affect the
observed power spectrum for a sampling pattern characterized by the
set of values $\SP$.  In this sense, our derivation is more along the
lines of \citet{VDK89}, explicitly showing what happens to the real
light curve in the process of obtaining the data.

\section{\textsc{Sampling Effects in The Power Spectrum}}
\label{sec:sampling}

From the theoretical point of view, it is common to model and
characterize the spectral content of a variable stochastic process by
specifying the underlying (or model) power spectrum, $\PM$, as a
continuous function of the frequency $\nu$. This spectrum can be
thought of as associated with a process, $C(t)$, that varies over time
in a continuous way.  In the context of X-ray variability, $C(t)$
could be given by the ``underlying'' (as opposed to observed)
continuous and infinite X-ray light curve.  In principle, complete
knowledge of the observable $C(t)$ implies complete knowledge of the
underlying spectrum, $\PM$, and vice versa. Therefore, if we had
unlimited access to the X-ray light curve, we could reconstruct the
underlying power spectrum accurately for every frequency.

In practice, however, we do not have access to the continuous light
curve, $C(t)$, but rather to a finite and discrete set of values
$\{C_{\rm o}(t_n)\}$ that results from observing the source for a
period of time $\TO$ every $t_n\equiv n\TS$, where $\TS$ is the
inverse of the sampling rate and $n=0, ..., N-1$ (without loss of
generality we will assume here that $N\equiv\TO/\TS$ is even).  In
order to work with reasonably good signal-to-noise ratios, the data
are usually binned for an interval of time $\TB \le \TS$ (objects are
said to be monitored ``continuously'' when $\TS \equiv \TB$). In what
follows we will refer to the set of values $\{\TO, \TS, \TB\}$ as the
``sampling pattern''.

The finite and discrete nature of the observations not only dictates
that we only have access to a finite and discrete set of values of the
observed power spectrum, i.e., $\{P_{\rm o}(\nu_k)\}$, with $\nu_k =
k/\TO$ and ${k=-N/2-1, ..., N/2}$, but also that, through the effects
of red noise leakage and aliasing \citep{Deeming75, VDK89}, these
values do not coincide with the values of the underlying (model) power
spectrum at those frequencies (i.e., $\{P_{\rm o}(\nu_k)\} \ne
\{P(\nu_k)\}$). Moreover, because of the stochastic nature of
the underlying process, the observed values $\{P_{\rm o}(\nu_k)\}$
differ, in general, from observation to observation. For a stationary
process, however, the average taken over different observations (with
the same sampling pattern), is well defined and it can be used as a
consistent estimator of the underlying power spectrum
\citep{Priestley89, Press92}.  The aim of this section is to find an
equation that relates the continuous underlying power spectrum, $\PM$,
with the set of values $\{\langle P_{\rm o}(\nu_k) \rangle \}$
obtained by averaging several observations with the same sampling
pattern.

Because our goal is to find an equation that involves the model power
spectrum, $\PM$, we need to work with continuous functions. In order
to do so, we describe the set of observed values $\{C_{\rm o}(t_n)\}$
in terms of the continuous variable $t$, representing the time, as
follows
\begin{eqnarray}
\label{eq:cobs_t}
C_{\rm o}(t) &=&  C_{bin}(t; \TB) \,W_{fin}(t; \TO, \TS) ~.
\end{eqnarray}
In this expression,
\begin{eqnarray}
\label{eq:cbin_t}
C_{bin}(t; \TB)= \frac{1}{\TB}  \int_{-\infty}^{\infty}  C(t') \, W(t-t'; \TB) \,dt'
\end{eqnarray}
is the ``continuous binned light curve'',
\begin{eqnarray}
\label{eq:comb_t_fin}
W_{fin}(t; \TO, \TS) =   W(t; \TO) \, \TS \!\!\!\!\sum_{n=-\infty}^{\infty}
\!\!\!\delta(t-n\TS)
\end{eqnarray}
is the ``finite comb function'', and  $W(t, \Delta t)$ is the boxcar
function of width $\Delta t$ given by
\begin{displaymath}
\label{eq:box_car_t}
W(t, \Delta t) = \left\{ \begin{array}{ll} 1 & \textrm{if} -\Delta t/2
\le t \le \Delta  t/2 \\ 0 & \textrm{otherwise} \\
\end{array} \right. ~.
\end{displaymath}

We can best interpret the right hand side of equation
(\ref{eq:cobs_t}) piecewise. The convolution of the continuous light
curve $C(t')$ with a boxcar function of width $\TB$ retrieves a new
continuous function of time, $C_{bin}(t;\TB)$. At any given time $t$,
the value $C_{bin}(t;\TB)$ is the average number of counts received in
the time interval $|t-t'|\le \TB$.  The finite series of equidistant
delta functions ensures that the observed light curve is not zero only
when we are sampling (i.e., when $t$ is some multiple of $\TS$ inside
the range of observation $[-\TO/2, \TO/2]$)\footnote{In order to have
  statistically independent measurements, we must have $\TS \ge \TB$,
  while, in order to minimize aliasing effects, we would like to have
  $\TS=\TB$. In practice, however, with the exception of short-term
  observations, the sampling time employed in AGN observations is much
  larger than the binning time, i.e., $\TS \gg \TB$.}.  Equation
(\ref{eq:cobs_t}) can be seen as the mathematical representation of
the process to which we subject the real (continuous) light curve when
we observe it.  Defined in this way, $C_{\rm o}(t)$ is a function of a
continuous variable, it is zero for all $t \ne t_n$, and its integral
over time is equal to the total number of counts detected during the
the entire period of observation $\TO$.

In order to understand how the observed values $\{\langle P_{\rm
  o}(\nu_k) \rangle \}$ differ from the underlying power spectrum due
to sampling effects, we need to calculate the Fourier Transform of the
light curve affected by the finite and discrete nature of the
observations, $C_{\rm o}(t)$, given by equation (\ref{eq:cobs_t}).  By
virtue of the convolution theorem, its Fourier Transform, $\tilde
C_{\rm o}(\nu)$, is given by
\begin{eqnarray}
\label{eq:tilde_cobs}
\tilde C_{\rm o}(\nu) = \! \int_{-\infty}^{\infty}  \!\!\!
\tilde C_{bin}(\nu';\TB) \, \tilde W_{fin}(\nu-\nu';\TO,\TS) \,d\nu' \,, \,\,\,\,\,
\end{eqnarray}
where $\tilde C_{bin}(\nu;\TB)$ and $\tilde W_{fin}(\nu;\TO,\TS)$ are
the Fourier Transforms of the continuous functions $C_{bin}(t; \TB)$
and $W_{fin}(t; \TO, \TS)$, respectively.  The Fourier Transform of
$C_{bin}(t;\TB)$ can be computed using the inverse of the convolution
theorem as
\begin{eqnarray}
\label{eq:tilde_cbin}
\tilde C_{bin}(\nu; \TB) = \frac{1}{\TB} \tilde W(\nu;\TB) \, \tilde
C(\nu) ~,
\end{eqnarray}
where $\tilde C(\nu)$ is the Fourier Transform of the continuous light
curve, $C(t)$, and
\begin{equation}
\label{eq:tilde_wtb}
\tilde W(\nu;\TB)= \frac{\sin(\pi \nu \TB)}{\pi \nu} = \TB \,
\sinc(\pi \nu \TB)
\end{equation}
is the Fourier Transform of the boxcar function of width $\TB$.

Using again the convolution theorem, we can write the Fourier
Transform of the finite comb function, $W_{fin}(t; \TO, \TS)$, 
defined in equation (\ref{eq:comb_t_fin}) as
\begin{equation}
\label{eq:tilde_wfin_conv}
\tilde W_{fin}(\nu;\TO,\TS)=\int_{-\infty}^{\infty} \tilde
\Sigma_\delta(\nu';\TS) \, \tilde W(\nu-\nu';\TO) \,d\nu' ~.
\end{equation}
Here, $\tilde W(\nu;\TO)$ is the Fourier Transform of the boxcar
function of width $\TO$, $W(t;\TO)$ (see eq.~[\ref{eq:tilde_wtb}]),
and $\tilde \Sigma_\delta(\nu;\TS)$ stands for the Fourier Transform
of ($\TS$ times) the ``infinite comb function'' at equidistant times
(see equation [\ref{eq:comb_t_fin}])
\begin{eqnarray}
  \tilde \Sigma_\delta(\nu;\TS) = 
  \!\!\sum_{m=-\infty}^{\infty}
  \!\!\delta\left(\nu-\frac{m}{\TS}\right)~,
\end{eqnarray}
which is an ``infinite comb function'' at equidistant frequencies
\citep[see, e.g.,][]{MF53}.  This expression allows us to carry out
the integral in equation (\ref{eq:tilde_wfin_conv}) which yields
\begin{eqnarray}
\label{eq:tilde_wfin}
\tilde W_{fin}(\nu;\TO,\TS)   = 
\TO
\!\! \sum_{m=-\infty}^{\infty}  \!\! \sinc[\pi(\nu-\nu_m)\TO] \,, \,\,\,\,\,
\end{eqnarray}
where we have defined the frequencies $\nu_m=m/\TS$.

Finally, using equations (\ref{eq:tilde_cbin}), (\ref{eq:tilde_wtb}),
and (\ref{eq:tilde_wfin}) we can write the Fourier Transform of the
function $C_{\rm o}(t)$ in equation (\ref{eq:tilde_cobs}) as
\begin{eqnarray}
\label{eq:Co_sum}
\tilde C_{\rm o}(\nu) = \!\!\sum_{m=-\infty}^{\infty} \!\!\tilde
C_{\rm WA}(\nu-\nu_m) ~,
\end{eqnarray}
where
\begin{eqnarray}
\label{eq:CWA_int}
\tilde C_{\rm WA}(\nu-\nu_m)  &=& \TO 
\int_{-\infty}^{\infty} \!\! \sinc(\pi \nu' \TB)  \, \tilde C(\nu')
\nonumber \\ &\times&\sinc[\pi(\nu-\nu'-\nu_m)\TO]  \,d\nu' ~,
\end{eqnarray}
can be interpreted as the ``windowed'' and ``aliased'' continuous
Fourier Transform associated with the real light curve $C(t)$.
Because of sampling effects, $\tilde C_{\rm o}(\nu)$ is a periodic
function of period $1/\TS$ (i.e., twice the Nyquist frequency,
$\nu_{\rm Nyq}\equiv1/2\TS$) and thus the frequency $\nu$ must be
chosen to vary over a frequency range spanning at most $1/\TS$. It is
customary to choose this range as $[-\nu_{\rm Nyq}, \nu_{\rm
  Nyq}]$. Even when this is the case, however, frequencies outside
this range will influence the value $\tilde C_{\rm o}(\nu)$ inside
$[-\nu_{\rm Nyq}, \nu_{\rm Nyq}]$.  When evaluated at the
Fourier frequencies $\nu_k = k/\TO$, the set $\{\tilde C_{\rm
  o}(\nu_k) \}$ is related to the \emph{observed} set of discrete
Fourier coefficients $\{c_k\}$ that we would have obtained by taking
the discrete Fourier Transform of the set of values $\{C_{\rm
  o}(t_n)\}$ via
\begin{eqnarray} 
\tilde C_{\rm o}(\nu_k) = \TO c_k \equiv \frac{\TO}{N}\sum_{n=0}^{N-1}
C_{\rm o}(t_n) ~e^{-2\pi i \nu_k t_n} ~.
\end{eqnarray}
The set of equations (\ref{eq:Co_sum}) and (\ref{eq:CWA_int}) show
explicitly how the discrete set of values $\{\tilde C_{\rm
  o}(\nu_k)\}$ is related to the \emph{underlying} spectral content of
the variable process, $\tilde C(\nu)$.

The relationship between the power spectrum affected by sampling
effects and the Fourier Transform of any given observed light curve is
given by $\PO = |\tilde C_{\rm o}(\nu)|^2 /\TO^2$. Therefore, using
equation (\ref{eq:Co_sum}), we can write
\begin{eqnarray}
\label{eq:P0_mk}
P_{\rm o}(\nu) =  \frac{1}{\TO^2} \sum_{m,m'=-\infty}^{\infty}  
\!\!\!\!\! \tilde C_{\rm WA}(\nu-\nu_m) \, \tilde C_{\rm WA}^{*}(\nu-\nu_{m'}) \,,  \,\,\,
\end{eqnarray}
where $\tilde C^{*}_{\rm WA}(\nu)$ stands for the complex conjugate of
$\tilde C_{\rm WA}(\nu)$.  However, when the underlying process is
stochastic, the variance associated with this estimator obtained from
any given observation is $100\%$ \citep{Priestley89, Press92}. In
practice, a consistent estimator of the underlying power spectrum is
obtained by considering the average value of the observed power
spectra obtained from many observations with the same sampling
pattern, $\SP$. For a stochastic process, the average over
observations of the cross-terms, i.e., $m' \ne m$, in equation
(\ref{eq:P0_mk}) vanishes \citep{Priestley89}, so we can write it as
\begin{eqnarray}
\label{eq:P0_mm}
\langle P_{\rm o}(\nu) \rangle = \frac{1}{\TO^2}
\sum_{m=-\infty}^{\infty} \langle |\tilde C_{\rm WA}(\nu-\nu_m)|^2 \rangle
~.
\end{eqnarray}
Using equation (\ref{eq:CWA_int}), the  average over observations
becomes
\begin{eqnarray}
\label{eq:psdo_sum_m}
\langle \PO \rangle &=& 
\!\!\sum_{m=-\infty}^{\infty} \int_{-\infty}^{\infty}
\int_{-\infty}^{\infty} \langle \tilde C(\nu') \tilde C^*(\nu'') \rangle \\ 
&\times& \sinc(\pi\nu'\,\TB) \, \sinc[\pi(\nu-\nu'-\nu_m) \,\TO] \,\, d\nu' \nonumber  \\
&\times& \sinc(\pi\nu''\TB) \sinc[\pi(\nu-\nu''-\nu_m)\TO] \, d\nu''
~. \nonumber
\end{eqnarray}

Furthermore, if the stochastic stationary process is ergodic, we can
evaluate the average, $\langle \tilde C(\nu') \,\tilde C^*(\nu'')
\rangle$, as an average over the ensemble of realizations $\{ \tilde
C(\nu)\}$.  In Appendix \ref{app:average} we provide a brief
demonstration showing that, in this case,
\begin{eqnarray}
\label{eq:CC_Pdelta}
\langle \tilde C(\nu') \,\tilde C^*(\nu'') \rangle =  P(\nu')
~\delta(\nu' - \nu'') ~.
\end{eqnarray}
In this way, we can carry out one of the integrals in equation
(\ref{eq:psdo_sum_m}) to obtain the relationship between the
underlying power spectrum, $\PM$, and the estimator obtained from
averaging over observations, $\langle \PO\rangle$, as
\begin{eqnarray} 
\label{eq:psdo_psd}
\langle \PO \rangle &=&   \int_{-\infty}^{\infty} \!\!
 P(\nu') \,\tilde \mathcal{W}\left(\nu, \nu'; \TO,\TS,\TB\right)
 \,d\nu' \,, \,\,\,
\end{eqnarray}
where the function
\begin{eqnarray} 
\label{eq:window_all}
\tilde  \mathcal{W}\left(\nu, \nu'; \TO,\TB,\TS\right) &=& 
\,\, 
\sinc^2(\pi\nu'\TB)  \\
&\times& \!\!\sum_{m=-\infty}^{\infty}
\!\!\sinc^2[\pi(\nu-\nu'-\nu_m)\TO] \nonumber 
\end{eqnarray}
is the ``window function'' that accounts for the effects of binning,
sampling and finite observation time, all at once. Strictly speaking,
equation (\ref{eq:window_all}) cannot be considered as the convolution
of the underlying power spectrum, $\PM$, with the function $\tilde
\mathcal{W} \left(\nu, \nu';\TO,\TS,\TB\right)$ since the latter
cannot be written as a function of the difference $\nu-\nu'$ alone.

\begin{figure}[t]
\includegraphics[width=\columnwidth,trim=0 5 0 10]{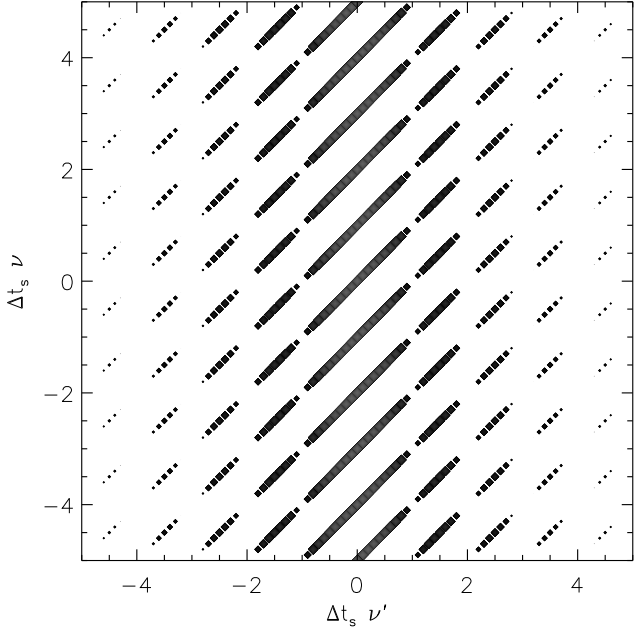}
\caption{Grey-scale representation of the ``window function'', $\tilde
  \mathcal{W}$ (see eq.~[\ref{eq:window_all}]), as a function of $\nu$
  and $\nu'$ for $\TS = \TB = 1024$s and $\TO=1024^2$s. Lighter shades
  of grey indicate higher values of $\tilde \mathcal{W}$. The function
  $\tilde \mathcal{W}$ reaches its maximum at $\nu'=0$ with subsidiary
  peaks of decaying strength on both sides.  As expected, the function
  $\tilde \mathcal{W}$ is periodic in $\nu$ with period $\TS$.}
\label{fig:window}
\end{figure}

As an aside, note that using the fact that the series of ``$\sinc$''
functions (weakly) converges to the Dirac delta, i.e.,
\begin{equation}
\lim_{\eta \rightarrow \infty} \eta \, \sinc^2(\pi \eta \nu) =
\delta(\nu) ~,
\end{equation}
it is not hard to show that, as expected,  in the limit of an infinite
and continuous observation the average of the observed power spectra tends
to the underlying power spectrum, i.e.,
\begin{equation}
\label{eq:continuous_limit}
\lim_{\TO \rightarrow \infty} \left(\lim_{\TS, \TB \rightarrow 0}
\TO \langle \PO \rangle \right)= \PM ~.
\end{equation}
Whenever a sampling pattern is applied, however, power at any given
frequency will ``leak'' to other frequencies, and vice versa. The
extent of this spectral leakage is fully characterized by the function
$\tilde \mathcal{W} \left(\nu, \nu';\TO,\TS,\TB\right)$ (see
Fig.~\ref{fig:window}).

\section{\textsc{Lorentzian Models for AGN Variability}}
\label{sec:lorentzian models for AGN}

In the previous section we showed how the average power spectrum
inferred from a set of observations with the same sampling pattern
differs from the underlying power spectrum.  We can now write the
average fractional rms that we would measure from such a set of
observations as
\begin{eqnarray}
\label{eq:rms_sampling}
\langle \rmssq(\TO,\TS,\TB; M_{\rm BH}) \rangle &=& \nonumber \\ 2 \!\!
\sum_{k=1}^{N/2-1} \!\! \langle P_{\rm o}(\nu_k; M_{\rm BH})\rangle &+&
\langle P_{\rm o}(\nu_{\rm Nyq}; M_{\rm BH})\rangle ~,
\end{eqnarray}
where the function $\langle \PO \rangle$ is the power spectrum
affected by sampling effects and is related to the underlying power
spectrum, $\PM$, according to equations (\ref{eq:psdo_psd}) and
(\ref{eq:window_all}).  Equation (\ref{eq:rms_sampling}) (together
with eqs.~[\ref{eq:psdo_psd}] and [\ref{eq:window_all}]) must be at
the core of any theoretical model addressing the functional form of
the $\rmssq$-$\mbh$ correlation and considering the distorting effects
imprinted in the data by the particular sampling pattern used to
obtain the observed fractional rms \citep[see also,][where the case
$\TB=\TS$ is discussed]{ONP05}.

\subsection{Advantages of Lorentzian Models}
\label{sucbsec:lorentzian models}

Inspired by the similarities shared by the the global timing
properties of AGN and early observations of X-ray binaries, previous
studies involving models of broad-band noise AGN power spectra are
mostly based on underlying power spectra characterized by broken power
laws.  However, over the last decade, thanks to the timing
capabilities of $\rxte$, it has been possible to study in great detail
the X-ray variability of many galactic sources. It is now clear that
the power spectra of X-ray binaries exhibit a rich morphology which,
in general, cannot be adequately described in terms of broken power
laws.

In many cases, it has been possible to obtain very good fits to the
observed power spectra (of both neutron stars and black holes) using a
small set of Lorentzians. It is important to point out that, whenever
checked against each other, Lorentzian models usually yield better
fits than power laws. Even when the fits are similarly good, the use
of Lorentzians allows us to ``follow'' over time the different
characteristic frequencies appearing in the power spectrum. This has
been key to uncovering several correlations among the different
broad-band noise components in X-ray binaries \citep{PBK99, BPK02}.

A typical Lorentzian model for the underlying power spectrum of an X-ray
binary can be written as
\begin{equation}
\label{eq:lorentzian_model}
P(\nu) 
= \sum_{l=1}^{N_L} L_l(\nu)
= \sum_{l=1}^{N_L}
 \frac{r_l^2}{\pi} ~\frac{\Delta_l}{\Delta^2_l + (\nu-\nu^0_l)^2} ~,
\end{equation}
where $N_L$ is the number of components, and $r_l$, $\Delta_l$, and
$\nu^0_l$ stand for the fractional rms, the width, and the centroid
frequency corresponding to each Lorentzian.  Defined in this way, the
power spectrum is ``two sided'' and therefore
\begin{equation}
r_l^2 =  \int_{-\infty}^{\infty} L_l(\nu) \,d\nu  \, .
\end{equation}

Although the data currently available for AGN might not allow yet for
a clear distinction between power-law models and Lorentzian models,
the underlying assumption of similar X-ray variability properties
between AGN and X-ray binaries advocates for the development of a
common phenomenological framework to study both kind of sources on the
same footing. Motivated by this, we propose to use a small set of
Lorentzians to model AGN power spectra.

\subsection{Sampling Effects in Lorentzian Models}
\label{sucbsec:sampling lorentzian models}

\begin{figure}[t]
\includegraphics[width=\columnwidth,trim=0 5 0 10]{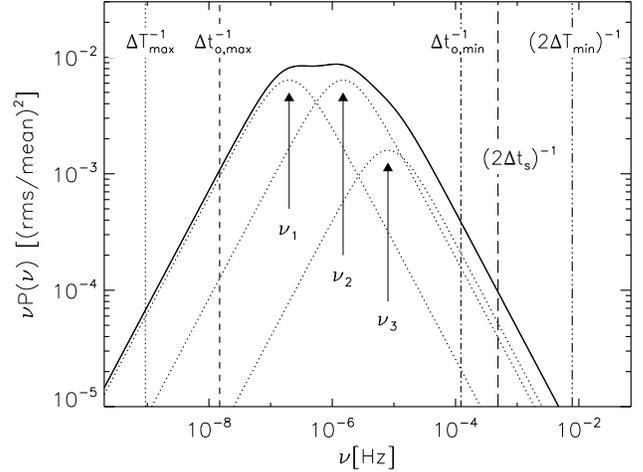}
\caption{The underlying power spectrum for a hypothetical AGN
  (\emph{solid line}) modeled as a sum of three broad Lorentzians
  (\emph{dotted lines}). The arrows indicate the positions of the peak
  frequencies (see Tab.~\ref{tab:lorentzians}). \emph{Vertical lines}
  indicate the frequencies corresponding to the different timescales
  involved in the numerical experiment performed in \S
  \ref{sec:contrast}.}
\label{fig:model_psd}
\end{figure}

Broad-band noise components are usually characterized by ratios
$\Delta_l/\nu^0_l \gg 1$ and, in most cases, it is a good
approximation to set $\nu^0_l=0$ in equation
(\ref{eq:lorentzian_model}) and describe the band-limited noise as a
sum of zero-centered Lorentzians \citep{BPK02}.  Let us then assume an
underlying broad-band noise power spectrum that can be described by a
set of $N_L$ broad Lorentzians\footnote{If required, the present
  formalism, including the derivation in Appendix \ref{app:integral},
  can be generalized straightforwardly to include narrow Lorentzian
  features, like the ones used to describe quasi-periodic
  oscillations.},
\begin{equation}
\label{eq:lorentzian_sum}
P(\nu) = 
\frac{1}{\pi} \sum_{l=1}^{N_L} r_l^2 ~\frac{\nu_l}{\nu^2_l + \nu^2}  ~,
\end{equation}
where $\nu_l$ is the HWHM of the $l^{th}$ zero-centered Lorentzian but
also plays the role of a ``peak frequency'' in the $\nu P(\nu)$
vs. $\nu$ representation (see Fig.~\ref{fig:model_psd}).

With the normalization chosen in equation (\ref{eq:lorentzian_sum}), the
variance of the light curves associated with the underlying power
spectrum, $\PM$, is indeed the fractional rms. In other words, the
total fractional rms corresponding to the band-limited noise spectrum
is just
\begin{equation}
\rmssq = 
\int_{-\infty}^{\infty} \!\! P(\nu) \, d\nu =
\sum_{l=1}^{N_L} r^2_l ~.
\end{equation}
This choice, together with the fact that the fractional rms is
dimensionless, determines the units of all the quantities related to
the power spectrum. In particular, the underlying (continuous) power
spectrum has units of inverse frequency while its discrete counterpart
(affected by sampling effects) is dimensionless. Because of this, the
discrete values $P_{\rm o}(\nu_k)$ must be multiplied by the observing
time $\TO$ in order to be compared with the underlying values
$P(\nu_k)$. This is also the reason for which the factor $\TO$ appears
explicitly in eq.~(\ref{eq:continuous_limit}).

\begin{deluxetable}{ccc}[t]
\tablewidth{0.4\textwidth}
\tablecaption{Sample Lorentzian Model for the Power Spectrum
\label{tab:lorentzians}}
\tablehead{\colhead{$L_l$} & \colhead{$r_l$} & \colhead{$\nu_{l}$[Hz]}}  
\startdata
$L_1(\nu)$ & 0.2 & 2.0  $\times 10^{-7}$  \\  
$L_2(\nu)$ & 0.2 & 1.5  $\times 10^{-6}$  \\  
$L_3(\nu)$ & 0.1 & 8.0  $\times 10^{-6}$  \\
\vspace{-2mm}
\enddata
\tablecomments{Parameters defining the sample power spectrum as a sum
  of three zero-centered broad Lorentzians. The fractional rms and the
  ratios between centroid frequencies for each Lorentzian are similar
  to those observed in Cyg X-1 \citep{Pet03}, while the peak frequency
  of the lowest Lorentzian (i.e., $\nu_1$) is similar to the low
  frequency break observed in NGC 3783 \citep{MEV03}.}
\end{deluxetable}

Figure~\ref{fig:model_psd} shows a sample Lorentzian model for the
broad-band noise power spectrum of a hypothetical AGN (see also
Tab.~\ref{tab:lorentzians}).  In this case, the fractional rms and the
ratios between centroid frequencies for each Lorentzian are similar to
those observed in Cyg X-1 \citep{Pet03}, while the peak frequency of
the lowest Lorentzian (i.e., $\nu_1$) is similar to the low frequency
break of NGC 3783 \citep{MEV03}.

Using equations (\ref{eq:psdo_psd}) and (\ref{eq:window_all}) and the
model power spectrum, $\PM$, given by equation
(\ref{eq:lorentzian_sum}) we can write the average observed power
spectrum, $\langle \PO \rangle$, as
\begin{eqnarray}
\label{eq:psd_lorentzian_sum}
\langle \PO \rangle = 
\sum_{l=1}^{N_L} \sum_{m=-\infty}^{\infty}  I_{l,m}(\nu),
\end{eqnarray}
with
\begin{eqnarray}
I_{l,m}(\nu) &=& \frac{r_l^2 b_l}{\pi} \int_{-\infty}^{\infty}  \!\!
\sinc^2(\Delta x') \, \frac{\sinc^2[a_m(\nu)-x']}{b_l^2 + x'^2}
\,dx'~, \nonumber \\
\end{eqnarray}
where we have defined the dimensionless variables
\begin{eqnarray}
b_l = \pi \TO \nu_l &,& \quad   \Delta = \TB/\TO \,  , \nonumber \\ x'
= \pi \TO ~\nu' &,& \quad a_m(\nu) = \pi \TO (\nu -\nu_m) \,.
\nonumber
\end{eqnarray}
This integral can be computed analytically in the complex plane using
a slight modification of the residues theorem (see Appendix
\ref{app:integral} for the details). The result is given by
\begin{eqnarray}
\label{eq:Imj_analytic}
I_{l,m}(\nu) &=& \frac{r_l^2 b_l}{\Delta^2} \sum_{i=1}^6 A^i_{l,m}(\nu),
\\
A^1_{l,m}(\nu) &=&\frac{1-(1-\Delta)\cos(2\Delta a_m)}{2
  a_m^2(a_m^2+b_l^2)}  ~, \nonumber \\
A^2_{l,m}(\nu) &=&-\frac{\sin(2\Delta a_m)(b_l^2+2a_m^2)}{2
  a_m^3(a_m^2+b_l^2)^2} ~, \nonumber \\
A^3_{l,m}(\nu) &=& \frac{ (b_l^2-a_m^2)\cos(2a_m)}{2(a_m^2+b_l^2)^2
b_l^3} ~e^{-2b_l} \sinh^2(\Delta b_l)  ~,\nonumber \\
A^4_{l,m}(\nu) &=& - \frac{ a_m\sin(2a_m)}{(a_m^2+b_l^2)^2b_l^2}
~e^{-2b_l} \sinh^2(\Delta b_l)  ~, \nonumber \\
A^5_{l,m}(\nu) &=&  \frac{(b_l^2-a_m^2) }{2 (a_m^2+b_l^2)^2 b_l^3}
~e^{-\Delta b_l} \sinh(\Delta b_l) ~, \nonumber \\
A^6_{l,m}(\nu) &=&\frac{\Delta}{2 a_m^2 b_l^2}  ~,\nonumber 
\end{eqnarray}
where we have omitted the frequency dependence of the coefficients
$a_m(\nu)$ on the right hand sides of the expressions for
$A^i_{l,m}(\nu)$.  The symmetry property
\begin{equation}
A^i_{l,-m}(-\nu)=A^i_{l,m}(\nu)
\end{equation} 
reflects the fact that sampling effects do not affect the parity of
the observed power spectrum, i.e., $\langle P_{\rm o}(\nu) \rangle =
\langle P_{\rm o}(-\nu)\rangle$.

It is important to summarize here what we have accomplished with
equations (\ref{eq:psd_lorentzian_sum}) and (\ref{eq:Imj_analytic}).
We have reduced the problem of accounting for the sampling effects
caused by an arbitrary sampling pattern, $\SP$, in any broad-band
noise power spectrum that can be decomposed into a sum of
zero-centered Lorentzians, to evaluating a simple sum.  In practice,
the sum in equation (\ref{eq:psd_lorentzian_sum}) converges extremely
fast and it is necessary to add only a few (order of five) terms at
any given frequency. This makes the process of accounting for sampling
effects in Lorentzian models highly efficient if the aim is to run a
grid of models to contrast them against observations in order to
constrain the functional form of the underlying power spectrum. We
stress that, if the underlying power spectrum is given by equation
(\ref{eq:lorentzian_sum}), the average $\langle \PO \rangle$
calculated according to equations (\ref{eq:psd_lorentzian_sum}) and
(\ref{eq:Imj_analytic}) provides the value to which Monte Carlo
simulations, devised to account for the sampling effects imprinted by
the sampling pattern $\SP$, will converge. In the next section we show
that this is indeed the case.

\section{\textsc{Contrasting the Analytical Prediction to Monte Carlo Simulations}}
\label{sec:contrast}

The aim of this section is two-fold. First, we want to test how well
our analytical prediction is able to account for the distortions
suffered by the underlying power spectrum due to sampling effects.
Second, we would like to obtain an idea about how important of an
effect one is neglecting when calculating the rms variability using
equation (\ref{eq:rms_nosampling}) instead of (\ref{eq:rms_sampling})
for a sampling pattern typical of AGN observations.  In order to do
this, we performed a series of numerical experiments calculating
independently both sides of equation (\ref{eq:rms_sampling}) assuming
the model power spectrum characterized by the three broad Lorentzians
shown in Table~\ref{tab:lorentzians} (see also Fig.~\ref{fig:model_psd}).

In order to test the validity of equation (\ref{eq:rms_sampling}), we
need to provide a good representation of the `infinite'' and
``continuous'' light curve, $C(t)$, associated with the underlying
power spectrum, $P(\nu)$. Then we can segment, bin, and sample this
``real'' light curve according to the sampling pattern $\SP$, obtain a
set of time series $\{C_o(t_n)\}$, and calculate the rms amplitude,
$\rmssq$, associated with each sampled light curve as
\begin{eqnarray}
\label{eq:rms_per_lightcurve}
\rmssq = \frac{1}{N-1} \sum_{n=0}^{N-1} |C_0(t_n) -  \bar{C}_0|^2  \,,
\end{eqnarray}
where $N=\TO/\TS$ and $\bar{C_0}$ is the average corresponding to the
particular time series $\{C_o(t_n)\}$. The mean value in equation
(\ref{eq:rms_sampling}), $\langle \rmssq \rangle$, can then obtained
by averaging over several ``observed'' light curves with the same
sampling pattern.

Of course, a numerical representation of the ``real'' light curve
$C(t)$ will still be given by a discrete set of values $\{C(t_l)\}$,
much in the same way as we can only evaluate the underlying
(continuous) power spectrum at a finite number of frequencies
$\{P(\nu_k)\}$.  The way of doing this is to consider a minimum
frequency, $\nu_{\rm min}$, and a maximum frequency, $\nu_{\rm max}$,
such that the contributions to the rms variability from power outside
the range $[\nu_{\rm min}, \nu_{\rm max}]$ is negligible, i.e.,
\begin{equation}
\label{eq:rms_numin_numax}
\int_{0}^{\nu_{\rm min}} \PM \,d\nu \,,    \int_{\nu_{\rm
max}}^{\infty} \PM \,d\nu\, \ll   \int_{\nu_{\rm min}}^{\nu_{\rm max}}
\PM \,d\nu ~.
\end{equation}
In the case of the model under consideration this can be accomplished
by choosing $\nu_{\rm min}\ll \nu_1$ and $\nu_{\rm max}\gg \nu_3$.
Note that this is a much stronger condition than requiring that
$\nu_{\rm min} \ll 1/\TO$ and/or $\nu_{\rm max} \gg 1/2\TS$ for the
particular sampling pattern under consideration. These conditions over
$\nu_{\rm min}$ and $\nu_{\rm max}$ are necessary but not sufficient.

For all practical purposes, the time series $\{C(t_l)\}$ obtained in
this way can be considered as an accurate representation of $C(t)$ in
the sense that the power spectrum estimated from it will not be
noticeably affected by sampling effects.  Note that, the set of values
$\{P_{\rm o}(\nu_k)\}$ corresponds to one given realization of the
stochastic process and, therefore, in general, it does not coincide
with the values of the underlying power spectrum $\{P(\nu_k)\}$ at any
given frequency even in this case!  However, the lack of sampling
effects is noticed in the absence of the trends that would be
introduced by red noise leakage and/or aliasing inside the range
$[\nu_{\rm min}, \nu_{\rm max}]$.

In order to set up the numerical experiment, we generate the
``continuous'' light curve by setting $\nu_{\rm min} = 1/\Delta T_{\rm
  max}= 9.3 \times 10^{-10}$Hz (i.e., $\Delta T_{\rm max} =2^{30}$s or
roughly 36 yrs) and $\nu_{\rm max} = \nu_{\rm Nyq} = 1/(2\Delta T_{\rm
  min})=7.8 \times 10^{-3}$Hz ($\Delta T_{\rm min}=64$s).  This yields
$N\equiv \Delta T_{\rm max}/\Delta T_{\rm min}=2^{24} \simeq 10^7$
initial ``data'' points.  The frequencies corresponding to $\nu_{\rm
  min}$ and $\nu_{\rm max}$ are indicated with vertical lines
(\emph{dotted} and \emph{triple-dot-dashed line}, respectively) in
Figure~\ref{fig:model_psd}. The conditions stated in equation
(\ref{eq:rms_numin_numax}) are clearly satisfied.

\begin{figure}[t]
\includegraphics[width=\columnwidth,trim=0 5 0 10]{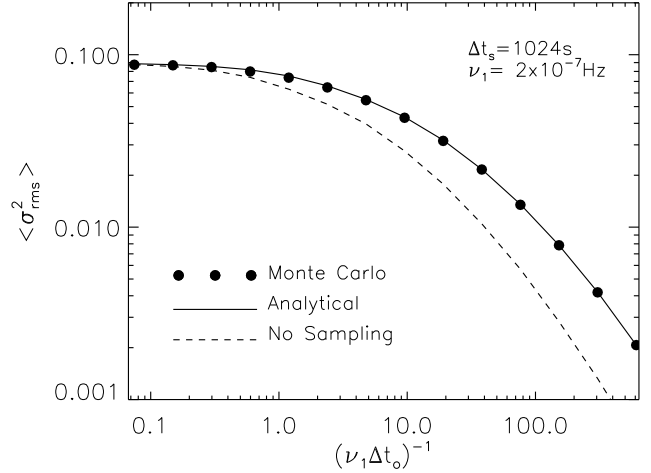}
\caption{The square of the fractional rms, $\rmssq$, associated with
  the model power spectrum, $\PM$, as a function of the
  observation time $\TO$ calculated in three different ways.
  \emph{Filled circles}: results obtained from Monte Carlo
  simulations.  \emph{Solid line}: the analytical prediction given by
  the right hand side of eq.~(\ref{eq:rms_sampling}).  \emph{Dashed
    line}: results obtained by neglecting sampling effects, i.e.,
  using eq.~(\ref{eq:rms_nosampling}). In this case, the binning time
  is equal to the sampling time.}
\label{fig:rms_tobs}
\end{figure}

Under our current set of assumptions, we can generate the time series
$\{C(t_l)\}$, with $t_l=l\Delta T_{\rm min}$ and $l=0, ..., N-1$ from
the set $\{P(\nu_k)\}$, with $\nu_k = k/\Delta T_{\rm max}$ and
$k=-N/2+1,...,N/2$, as 
\begin{equation} 
C(t_l) = \!\!\sum_{k=N/2-1}^{N/2} \!\! c_k ~e^{2\pi i \nu_k t_l} ~,
\end{equation}
with
\begin{equation} 
\label{eq:ft_cont}
c_k =\sqrt{\frac{P(\nu_k)}{\Delta T_{\rm max}}} ~e^{2\pi i
\phi(\nu_k)} ~.
\end{equation}
In order to generate a true realization of the stochastic process with
underlying power spectrum $P(\nu)$, the amplitudes under the square
root in equation (\ref{eq:ft_cont}) must satisfy $P(\nu_k) =
P(-\nu_k)$ and be drawn from a $\chi^2$ distribution with two degrees
of freedom for all $\nu_k \ne \nu_{\rm Nyq}$ and from a $\chi^2$
distribution with one degree of freedom for $\nu_{\rm Nyq}$.
Alternatively, this can also be done by generating two normally
distributed random variables, multiplying them by $[P(\nu_k)/\Delta
T_{\rm max}]^{1/2}$, and using them to define the real and imaginary
parts of the discrete Fourier Transform given by the set
$\{c_k\}$\citep[see, e.g.,][for more details]{TK95}. Moreover, in
order to ensure that the generated light curve is real, the set of
random phases $\{\phi(\nu_k)\}$ must satisfy $\{\phi(-\nu_k)\} =
\{-\phi(\nu_k)\}$ for all $k\ne 0, N/2$ and $\phi(0) = 0$ or $\pi$, as
well as $\phi(\nu_{\rm Nyq}) = 0$ or $\pi$.

Note that, by construction, if we were to take the discrete Fourier
Transform of the set $\{C(t_l)\}$ and calculate $\{P(\nu_k)\} = \Delta
T_{\rm max} \{|c_k|^2\}$ then sampling effects would be negligible in
the range $[\nu_{\rm min}, \nu_{\rm max}]$.  For all practical
purposes then, the set $\{C(t_l)\}$ can be thought of as an accurate
representation of the ``infinite'' and ``continuous'' light curve
$C(t)$.

We can now define a sampling pattern, $\{\TO, \TS, \TB\}$, and do
everything we would do if we were recording a time series from the
continuous light curve $C(t)$.  In order to be able to compute an
average value for the rms variability we divided the original light
curve of duration $\Delta T_{\rm max}$ in 16 light curves of duration
$\Delta t_{\rm o, max} = 2^{26}$s. We also defined the duration of
each time bin as $\TB = 1024$s, as a representative value used in AGN
observations. Note that this value of $\TB$ will include $16$ discrete
points of the original time series $\{C(t_l)\}$.  To simplify the
discussion, in what follows, we will consider a continuous monitoring
campaign, i.e., $\TS=\TB$ (we consider a case in which $\TS > \TB$ in
the next section).

We calculated the fractional rms variability for each light curve of
duration $\Delta t_{\rm o, max}$ according to equation
(\ref{eq:rms_per_lightcurve}) and obtained the average value that we
took as the left hand side of equation (\ref{eq:rms_sampling}).  We
repeated this procedure defining an increasingly larger number of
segmented (independent) light curves of duration $\Delta t_{{\rm o}} =
2^{-n}\Delta t_{\rm o, max}$ with $n=0,...,13$.
Figure~\ref{fig:rms_tobs} shows the values of $\langle \rmssq \rangle$
as a function of the observing time\footnote{We have used the
dimensionless quantity $\nu_1\Delta t_{{\rm o}}$ as the independent
variable. In this case, sampling effects are not important for
$\nu_1\Delta t_{{\rm o}}\le 1$.}, $\Delta t_{{\rm o}}$.  The
analytical prediction (i.e., the right hand side of
eq.~[\ref{eq:rms_sampling}]) is shown with a solid line, the filled
circles represent the average values $\langle \rmssq \rangle$ obtained
from the light curves generated with the Monte Carlo simulations, and
the dashed line shows the value of $\rmssq$ as obtained using equation
(\ref{eq:rms_nosampling}).

\begin{deluxetable}{ccc}[t]
\tablewidth{0.35\textwidth}
\tablecaption{Sampling Effects on the Fractional RMS  
\label{tab:rms_comparison}}
\tablehead{\colhead{n} & \colhead{$\TO/\TS$} & \colhead{$\sigma^2_{\rm rms,WA}/\sigma^2_{\rm rms,NS}$}}
\startdata
\,\,\,3  &  8192     &     $\sim 10.00\,\%$   \\      
\,\,\,4  &  4096     &     $\sim 20.00\,\%$   \\      
\,\,\,9  &  128.0    &     $\sim 100.0\,\%$   \\      
12 &  16.00    &     $\sim200.0\,\%$   \\      
\vspace{-2mm} \enddata \tablecomments{Comparison between the
fractional rms obtained considering and neglecting sampling effects as
a function of the ratio between the observing and the sampling times.
The symbols $\sigma^2_{\rm rms,WA}$ and $\sigma^2_{\rm rms, NS}$
denote the fractional rms derived from the ``windowed'' and
``aliased'' power spectrum (eq.~[\ref{eq:rms_sampling}]), and the
value obtained when sampling effects are neglected
(eq.~[\ref{eq:rms_nosampling}]).}
\end{deluxetable}

It is important to stress that we have not adjusted the normalization
of any quantity when plotting the different variances in
Figure~\ref{fig:rms_tobs}. The convergence of the different fractional
amplitudes for large values of $\TO$ to the total rms amplitude of the
model (i.e., $\rmssq=0.09$) indicates that sampling effects become
negligible when the harmonic content of the observed light curve is a
good representation of the underlying power spectrum. This is in
agreement with equation (\ref{eq:continuous_limit}). As the
observation time decreases, the fractional rms, as calculated
according to equation (\ref{eq:rms_nosampling}), underestimates the
values obtained from either side of equation (\ref{eq:rms_sampling}),
i.e., the fractional rms obtained from either the analytical result or
the Monte Carlo simulations, by a factor of up to $200\%$ (see
Tab.~\ref{tab:rms_comparison} for details).

\begin{figure}[t]
\includegraphics[width=\columnwidth,trim=0 5 0 10]{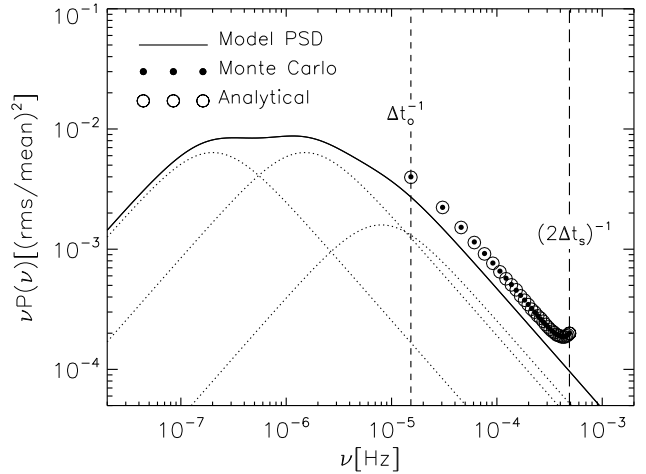}
\caption{Sampling distortions suffered by the model power spectrum
  defined in Tab.~\ref{tab:lorentzians} (\emph{solid line}), as
  obtained from calculating the (average) discrete power spectrum
  associated with the light curves generated via Monte Carlo
  simulations (\emph{filled circles}), and as predicted by our
  analytical calculation (\emph{open circles}).  
The sampling pattern is defined by $\TO=65536$s and
  $\TS=\TB=1024$s. Note that the discrete values $P_{\rm o}(\nu_k)$
  must be multiplied by $\TO$ in order to be compared with the
  underlying values $P(\nu_k)$.}
\label{fig:aliased_psd}
\end{figure}

In the previous numerical experiments, the fractional rms was obtained
directly from the light curves without the need of deriving the power
spectra associated with them. Figure~\ref{fig:aliased_psd} shows the
average power spectra derived from the Monte Carlo simulations
(\emph{filled circles}) for the case $\TS=\TB=1024$s and $\TO=65536$s
(corresponding to $n=10$) together with the corresponding analytical
result (\emph{open circles}) derived using equations
(\ref{eq:psd_lorentzian_sum})-(\ref{eq:Imj_analytic}). The distorting
effects suffered by the power spectrum derived from the finite and
discrete light curves with respect to the underlying power spectrum
(\emph{solid line}) are evident. This remarkable agreement between the
averaged observed power spectra derived analytically and using the
simulated light curves is the reason behind the coincidence of the
associated fractional variability amplitudes in
Figure~\ref{fig:rms_tobs}.

As a corollary, we note that, by adopting a Lorentzian model for the
underlying power spectrum, the results from \S \ref{sucbsec:sampling
lorentzian models} can be used to test Monte Carlo simulations when
the frequencies are assumed to be uncorrelated.

\section{\textsc{Discussion}}
\label{sec:discussion}

This is the first paper in a series aimed to exploit the similarities
observed in the X-ray timing properties of AGN and X-ray binaries in
order to measure the masses of supermassive black holes using X-ray
variability.

The idea of using X-ray variability to obtain the masses of AGN by
comparing their timing properties to those of X-ray binaries is not
new \citep{Hayashida98, LY01, Czerny01, BZ03, P04, NPC04}.  A key
assumption necessary for this procedure to provide reliable results is
that the processes driving the variability, and therefore the
associated underlying power spectra, are similar. Although the data
currently available for some AGN is encouraging in this respect
\citep{UMP02, MEV03}, so far there have been no attempts of describing
the timing properties of both kind of objects on the same footing.
Motivated by the latest timing studies carried out on a number of
galactic black-hole candidates, we have proposed to model the
broad-band noise spectra of AGN in the same way that has been
successful with many X-ray binaries, i.e., as a sum of Lorentzian
components.

A particular concern when connecting theoretical models of broad-band
noise X-ray variability with observations is that sampling effects
(especially red noise leakage) are much more important for AGN than
for X-ray binaries. In the last few years it has become evident that
in order to derive reliable power-spectra estimates from the
observations it is imperative to account for these sampling effects
\citep{UMP02}. Less attention has been paid, however, to the effects
that sampling has on the connection between the underlying power
spectrum and the rms variability \citep[but see][]{ONP05}.

\begin{figure}[t]
\includegraphics[width=\columnwidth,trim=0 5 0 10]{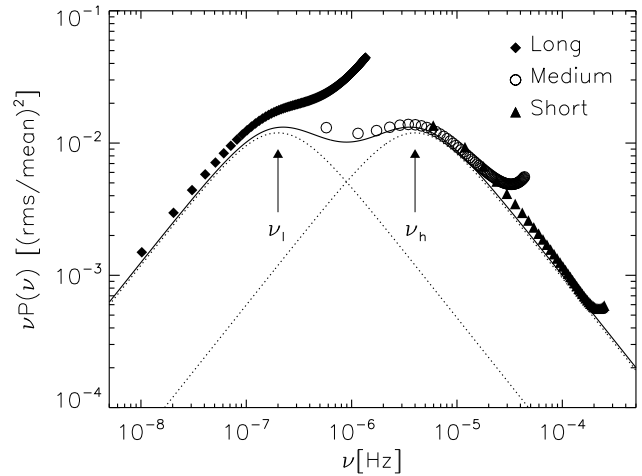}
\caption{Illustration of the distorting effects suffered by the
  underlying power spectrum, $P(\nu)$ (\emph{solid lines}), modeled
  with two Lorentzian components (\emph{dotted lines}, see
  eq.~[\ref{eq:p_mod_ngc3783}]) when it is ``observed'' according to
  different sampling patterns (see Tab.~\ref{tab:uneven_sampling}).
  The different symbols, \emph{filled diamonds}, \emph{open circles},
  and \emph{filled triangles}, indicate the ``windowed'' and
  ``aliased'' power spectra that would be derived from long-, medium-,
  and short-term observations respectively.}
\label{fig:uneven_sampling}
\end{figure}

In order to illustrate the importance of this latter issue, let us
assume a hypothetical Lorentzian model resembling the underlying power
spectrum obtained for NGC 3783.  \citet{MEV03} found that a reasonable
description of this power spectrum can be obtained by a double broken
power-law model with characteristic (``low'' and ``high'') break
frequencies given by $\nu_l=2.00 \times 10^{-7}$Hz and $\nu_h=3.98
\times 10^{-6}$Hz, and an amplitude $A \simeq 0.01$ (the corresponding
power-law slopes at low and intermediate frequencies were fixed to 0
and -1 respectively and the best fit at high frequencies is consistent
with a power-law of -2).  A Lorentzian model to describe this
underlying power spectrum could be obtained using two Lorentzian
components (\emph{dotted lines} in Fig.~\ref{fig:uneven_sampling}) as
follows
\begin{equation}
\label{eq:p_mod_ngc3783}
P(\nu) = \frac{r_1^2}{\pi} \frac{\nu_1}{\nu_1^2+\nu^2} +
\frac{r_2^2}{\pi} \frac{\nu_2}{\nu_2^2+\nu^2} ~,
\end{equation}
where we have denoted the characteristic frequencies as $\nu_1=\nu_l$
and $\nu_2=\nu_h$. We assume here that $r_1^2 = r_2^2 = 0.075$ so that
the peak amplitude of each Lorentzian in $\nu P(\nu)$ space is roughly
0.01.  As an aside, we note that the ratio of frequencies derived from
observations is such that these two Lorentzians (with the same
amplitude) lead naturally to a flat slope in the $\nu P(\nu)$
representation. This feature is observed in several X-ray binaries,
i.e., in many cases the frequencies and amplitudes of the different
broad Lorentzians are such that the overall shape of the (underlying)
power spectrum resembles a $1/\nu$ power law over a wide range in
frequencies \citep{BPK02}.

\begin{deluxetable}{lcccc}[t]
\tablewidth{0.45\textwidth}
\tablecaption{Typical Sampling Patterns for AGN
\label{tab:agn_sampling}}   \tablehead{Term Type &\colhead{$\TO$} &
\colhead{$\TS$} &  \colhead{$\TB$} & \colhead{$\sigma^2_{\rm rms,WA}/\sigma^2_{\rm rms,NS}$}}
\startdata   
Long     & 4.27$\times$266 d   &  4.27 d &  1024.8 s  & $\sim 78\,\%$  \\  
Medium   & 3.20$\times$151 h   &  3.20 h &  1152.0 s  & $\sim 32\,\%$  \\  
Short    & 2000$\times$84 s    &  2000 s &  2000.0 s  & $\sim 77\,\%$  \\  
\vspace{-2mm} 
\enddata 
\tablecomments{The sampling patterns for
short-term observations are typical of $\asca$, $\chandra$, and/or
$\xmm$. The values of $\SP$ for long-term observations are more
representative of $\rxte$ observations. The meaning of the symbols
$\sigma^2_{\rm rms,WA}$ and $\sigma^2_{\rm rms,NS}$ is the same as in
Tab.~\ref{tab:rms_comparison}.}
\label{tab:uneven_sampling}
\end{deluxetable}

Let us now suppose that we ``observe'' the model defined by equation
(\ref{eq:p_mod_ngc3783}) according to the different sampling patterns
involved in the various types (long-, medium- and short-term) of
monitoring campaigns designed to observe NGC 3783 (see
Tab.~\ref{tab:uneven_sampling}).  By calculating the rms variability
according to equation (\ref{eq:rms_nosampling}), ($\sigma^2_{\rm rms,
NS}$, i.e., not considering sampling effects), and equation
(\ref{eq:rms_sampling}), ($\sigma^2_{\rm rms, WA}$, i.e., when the rms
variability is calculated based on the ``windowed'' and ``aliased''
power spectrum) we can study how important are the consequences of
neglecting sampling effects.  The last column in
Table~\ref{tab:uneven_sampling} shows the ratio of these two variances
for the different types of sampling patterns.  For the adopted power
spectrum, neglecting sampling effects can lead us to underestimate the
observed value of the rms variability by up to $80\%$.  These
differences can also be understood by looking at the sampling effects
directly in the power spectrum.  Figure~\ref{fig:uneven_sampling}
shows the ``observed'' power spectra that we would obtain from
monitoring a source with an underlying power spectrum given by
equation (\ref{eq:p_mod_ngc3783}), according to the sampling patterns
listed in Table~\ref{tab:uneven_sampling}.

It is not straightforward to see how these differences in the
variances will affect the mass estimates derived from a theoretical
model of the $\rmssq$-$\mbh$ correlation. This will depend, of course,
upon the functional form of the $\rmssq$-$\mbh$ correlation itself. We
note here that a difference in rms variability will likely be
amplified when translated into a difference in mass.  This is because
the observed fractional rms is a rather flat function of black-hole
mass.  Indeed, for a large number of AGN the fractional rms is known
to vary between $30\%$ and $40\%$ in long-term observations and
between $2\%$ and $30\%$ in short-term observations, over almost four
orders of magnitude in black-hole mass \citep{MEV03}.

This flattening of the observed $\rmssq$-$\mbh$ correlation for larger
observation times suggest that, for the sake of deriving masses, short
observations should be favored over long term monitoring campaigns.
This is, of course, when sampling effects (in particular red noise
leakage) will affect the most the connection between the model power
spectrum and the observed fractional rms variability.  For this
reason, it is vital to incorporate sampling effects in theoretical
models that aim to estimate black-hole masses via the $\rmssq$-$\mbh$
correlation if they rely on a parametrization of the underlying AGN
power spectrum in terms of $\mbh$.

It is evident from Figure~\ref{fig:uneven_sampling} that the effects
of spectral leakage have to be considered when (continuous) model
power spectra are to be compared against (average) power spectra
obtained from observed time series.  State of the art techniques to
``convolve'' the models with the observation sampling patterns
currently involve sophisticated Monte Carlo methods.  These
simulations are crucial once the ``best fit model'' has been found in
order to assess the confidence levels associated with the best fit
parameters.  The high numerical costs associated with this type of
simulations, however, usually rules in favor of mathematical models
that can be described with a low number of free parameters which might
not allow enough freedom to properly describe the data.

The proposed Lorentzian models for broad-band noise variability offer
a tremendous advantage in this sense.  In the measure that the
available data allows it, the gain in computational speed obtained by
using analytical expressions to account for sampling effects could be
used to explore parameter spaces of higher dimensions. This will play
in favor of more realistic mathematical models as it is the case for
X-ray binaries.

\acknowledgments{I thank Dimitrios Psaltis for valuable discussions
  and his encouragement and support throughout this work. I am
  grateful to Tomaso Belloni for reading the manuscript and providing
  useful comments and to the referee, Michiel van der Klis, for
  detailed comments that helped improve this manuscript.  I also
  acknowledge the hospitality of the Astronomical Institute ``Anton
  Pannekoek'' during part of this study. This work was partially
  supported by NASA grant NAG-513374.}


\begin{appendix}

\section{\textsc{Average Over Realizations}}
\label{app:average}

In \S \ref{sec:sampling} we stated that the average over ensembles of
the quantity $\langle \tilde C(\nu') \tilde C^{*}(\nu'')\rangle$ was
related to the underlying power spectrum of a stationary stochastic
process via equation (\ref{eq:CC_Pdelta}).  Here, we provide a brief
demonstration of this statement.

By definition, the average $\langle \tilde C(\nu') \tilde C^*(\nu'')
\rangle$ can be written in terms of the corresponding average in the
time domain as
\begin{equation}
  \langle \tilde C(\nu') \tilde C^*(\nu'') \rangle =
  \int_{-\infty}^{\infty} \int_{-\infty}^{\infty} \langle C(t') C^*(t'')
  \rangle  ~e^{-i2\pi\nu' t'}  ~e^{i2\pi\nu'' t''} dt' dt'' ~,
\end{equation}
where $C(t)$ corresponds to the inverse Fourier Transform of $\tilde
C(\nu)$.  For a continuous process with zero mean, the quantity
$\langle C(t') C^*(t'')\rangle$ is the correlation function $\xi(t',
t'')$.  Furthermore, if the process is stationary, the expectation
value $\langle C(t') C^*(t'') \rangle$ can only be a function of the
time difference $t'-t''$ and, therefore,
\begin{equation}
\langle \tilde C(\nu') \tilde C^*(\nu'') \rangle =
\int_{-\infty}^{\infty} \int_{-\infty}^{\infty} \xi(t'-t'')
~e^{-i2\pi\nu' t'}  ~e^{i2\pi\nu'' t''} ~dt' dt''~.
\end{equation}
Defining  $\tau = t'-t''$ and $\tilde \tau = t'$, we can write
\begin{eqnarray}
\langle \tilde C(\nu') \tilde C^*(\nu'') \rangle &=&
\int_{-\infty}^{\infty} \int_{-\infty}^{\infty} \xi(\tau)
e^{-i2\pi\nu' \tilde \tau}  ~e^{i2\pi\nu'' (\tilde \tau -  \tau)}
d\tau d\tilde \tau ~, \\ &=& \int_{-\infty}^{\infty}  \xi(\tau)
e^{-i2\pi\nu''\tau}  ~d\tau \int_{-\infty}^{\infty} e^{-i2\pi (\nu' -
\nu'') \tilde \tau}  ~d\tilde \tau ~, \\  &=& F[\xi](\nu'')
~\delta(\nu' - \nu'')~,
\end{eqnarray}
where the quantity $F[\xi](\nu'')$ stands for the Fourier Transform of
the autocorrelation function, which is the underlying power spectrum,
$P(\nu'')$. In this way, we obtain the result quoted in equation
(\ref{eq:CC_Pdelta}), i.e., 
\begin{equation}
\langle \tilde C(\nu') \tilde C^*(\nu'') \rangle =  P(\nu'')
~\delta(\nu' - \nu'') \,.
\end{equation}

\section{\textsc{Contour Integral}}
\label{app:integral}

\begin{figure}[t]
\epsscale{0.6} \plotone{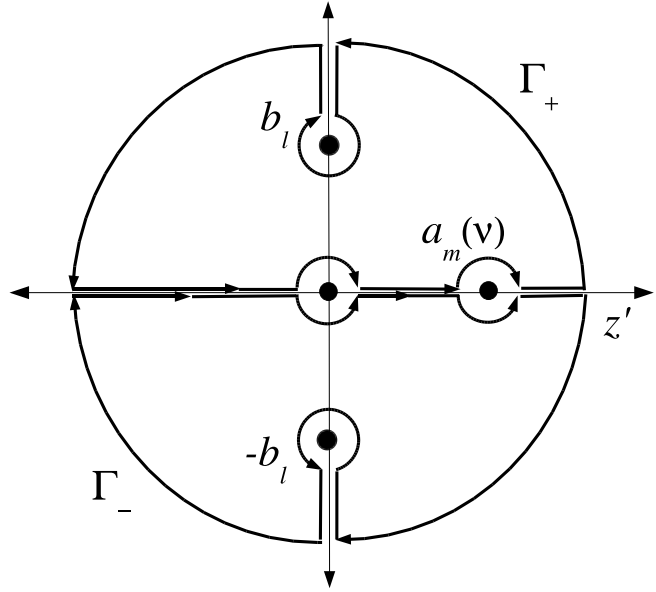}
\caption{The integration paths in the upper
($\Gamma_{+}$) and lower ($\Gamma_{-}$) halves of the complex plane
used to calculate the coefficients $I_{l,m}(\nu)$ in equation
(\ref{eq:I_mj_A1}). The functions $f_p^{\pm}$ defined in equation
(\ref{eq:f_pm}) decay exponentially over $\Gamma_{\pm}$ for ${\mathcal
Im}(z) \gtrless 0$, respectively, when the radii of the contours tend
to infinity.}
\label{fig:path}
\end{figure}

In \S \ref{sucbsec:sampling lorentzian models} we presented an
analytic expression for the coefficients $I_{l,m}(\nu)$ involved in
the calculation of the average power spectrum affected by sampling
effects according to the sampling pattern $\SP$. Here, we outline how
to obtain them.

The integral defining the coefficients $I_{l,m}(\nu)$, i.e., 
\begin{eqnarray}
\label{eq:I_mj_A1}
I_{l,m}(\nu) &=& \frac{r_l^2 b_l}{\pi} \int_{-\infty}^{\infty}
~\frac{\sin^2(\Delta x')}{(\Delta x')^2}
~\frac{\sin^2[a_m(\nu)-x']}{[a_m(\nu)-x']^2} ~\frac{dx'}{b_l^2 + x'^2}
~,
\end{eqnarray}
exists and is real for any non-zero value of $b_l$ and any finite
value of $\Delta$, and $a_m(\nu)$.  A convenient way to calculate its
value is to allow the variable $x'$ to be complex and compute
$I_{l,m}(\nu)$ as a contour integral in the complex plane by choosing
suitable integration paths.  The idea, as usual, is to find a contour
of integration that contains the real axis and close it conveniently
in such a way that the imaginary contribution vanishes when the real
part of the contour extends from $-\infty$ to $\infty$.  In order to
do so, it is necessary to rewrite the sines in terms of complex
exponentials that decay in the upper or lower half of the complex
plane depending on their sign. We can rewrite equation
(\ref{eq:I_mj_A1}) as
\begin{eqnarray}
\label{eq:I_mj_A2}
I_{l,m}(\nu) &=& \frac{r_l^2 b_l}{16\pi}\sum_{p=1}^{3} \sum_{\pm}
\int_{-\infty}^{\infty} \frac{f^{\pm}_p(z')}{(\Delta z')^2
[a_m(\nu)-z']^2} ~\frac{dz'}{b_l^2 + z'^2} ~.
\end{eqnarray}
where the functions $f_p^{\pm}$ for $p=1,2,3$ are given by
\begin{eqnarray}
\label{eq:f_pm}
f^{\pm}_1(z) &=& (e^{2i\Delta z}-1)  ~e^{\pm i[2(z-a_m)]} ~,   \nonumber
\\   f^{\pm}_2(z) &=& (e^{-2i\Delta z}-1) ~e^{\pm i[2(z-a_m)]} ~,
\nonumber \\ f^{\pm}_3(z) &=& 2 (1-e^{\pm2i\Delta z})~.
\end{eqnarray}
Note that because $\Delta = \TB/\TO <1$, the functions $f_p^{+}$ and
$f_p^{-}$, for $p=1,2,3$, decay exponentially in the upper and lower
halves of the complex plane respectively.  A minor complication arises
because the integrand presents (simple) poles on the real axis, i.e.,
at $z_o=0$ and $z_m=a_m(\nu)$, so some care is required when choosing
the contours of integration in either half. Figure~\ref{fig:path}
shows a possible way of choosing two closed paths. The contours
$\Gamma_{\pm}$ are adequate to calculate the integrals containing the
functions $f_p^{\pm}$, respectively.  Using a slight modification of
the residues theorem \citep[see, e.g.,][]{Marsden&Hoffman3rd} we can
calculate the integrals in equation (\ref{eq:I_mj_A2}) to obtain
\begin{eqnarray}
\label{eq:I_mj_A3}
I_{l,m}(\nu) &=& \frac{r_l^2 b_l}{16\pi}\sum_{p=1}^{3} \sum_{\pm} \pm
2\pi i {\mathcal Res}(f_p^{\pm}, \pm b_l)  \pm \pi i {\mathcal
  Res}(f_p^{\pm}, 0) \pm \pi i {\mathcal Res}[f_p^{\pm}, a_m(\nu)] ~,
\end{eqnarray}
where ${\mathcal Res}(f,z)$ stands for the residue of the function $f$
evaluated at the (simple) pole $z$. Note that the signs in this
equation properly take care of the negative orientation of the contour
$\Gamma_{-}$. After some lengthy, but otherwise straightforward,
algebra to regroup similar terms, equation (\ref{eq:I_mj_A3}) yields
the result quoted in equation (\ref{eq:Imj_analytic}).

\end{appendix}

\end{document}